\documentclass[aps,twocolumn,pra,superscriptaddress,nofootinbib,10pt]{revtex4-1}
\usepackage[]{amsmath}
\usepackage{graphics}
\usepackage{amssymb}
\usepackage{graphicx}
\usepackage[usenames,dvipsnames,svgnames]{xcolor}
\usepackage{lipsum}
\usepackage{soul}

\newcommand{\be}{\begin{equation}}
\newcommand{\ee}[1]{\label{#1} \end{equation}}

\def\f12{\frac{1}{2}}

\newcommand\blfootnote[1]{%
  \begingroup
  \renewcommand\thefootnote{}\footnote{#1}%
  \addtocounter{footnote}{-1}%
  \endgroup
}

\begin{document}
\title{Gravitational wave detection with optical lattice atomic clocks}

\author{S. Kolkowitz$^{*,1}$,
I. Pikovski$^{2,3}$,
N. Langellier$^{2}$,
M.D. Lukin$^{2}$,
R.L. Walsworth$^{2,4}$,
J. Ye$^{\dagger,}$}

\affiliation{JILA, NIST and University of Colorado, 440 UCB, Boulder, Colorado 80309, USA,\\
$^{2}$Department of Physics, Harvard University, Cambridge, Massachusetts 02138, USA,\\
$^{3}$ITAMP, Harvard-Smithsonian Center for Astrophysics, Cambridge, Massachusetts 02138, USA \\
$^{4}$Harvard-Smithsonian Center for Astrophysics and Center for Brain Science, Cambridge, Massachusetts 02138, USA}

\begin{abstract}
We propose a space-based gravitational wave detector consisting of two spatially separated, drag-free satellites sharing ultra-stable optical laser light over a single baseline. Each satellite contains an optical lattice atomic clock, which serves as a sensitive, narrowband detector of the local frequency of the shared laser light. A synchronized two-clock comparison between the satellites will be sensitive to the effective Doppler shifts induced by incident gravitational waves (GWs) at a level competitive with other proposed space-based GW detectors, while providing complementary features. The detected signal is a differential frequency shift of the shared laser light due to the relative velocity of the satellites, and the detection window can be tuned through the control sequence applied to the atoms' internal states. This scheme enables the detection of GWs from continuous, spectrally narrow sources, such as compact binary inspirals, with frequencies ranging from $\sim3$~mHz - 10 Hz without loss of sensitivity, thereby bridging the detection gap between space-based and terrestrial optical interferometric GW detectors. Our proposed GW detector employs just two satellites, is compatible with integration with an optical interferometric detector, and requires only realistic improvements to existing ground-based clock and laser technologies.
\end{abstract}

\blfootnote{$^{*}$shimonk@jila.colorado.edu\\
$^{\dagger}$ye@jila.colorado.edu }

\date{\today}
\maketitle

The first direct detections of gravitational waves (GWs) by the Laser Interferometer Gravitational-Wave Observatory (LIGO) \cite{LIGOdetection,LIGOdetection2} heralds the dawn of a new era of astrophysics. The culmination of a century-long search \cite{LIGOdetection,LIGOdetection2,LIGO,TorsionBar,SeismoGW,detweiler1979pulsar,GPS,Ulysses,Apollo,DopplerReview}, GW detection is now emerging as a new tool with which to study the universe, illuminating previously invisible astrophysical phenomena. In parallel, the developments of laser cooling and the laser frequency comb have given rise to optical atomic clocks with accuracies and stabilities at the 10$^{-18}$ level \cite{ClockSr,ClockYb,Katori2015,ClockBetter,PTBion}. As clock precision continues to improve, there is growing interest in the prospect of using optical atomic clocks for GW detection \cite{DopplerReview,LoebGWclocks,Vutha}. In this work we outline a proposal for a new GW detector based on Doppler shift measurements between
two spacecraft containing optical lattice atomic clocks linked over a single optical baseline.
This detector offers broad tunability of narrowband sensitivity in the mHz - Hz frequency range. As GW astronomy matures, such a detector can therefore serve as a different type of observatory for gravitational waves that can be complementary to existing concepts, much like there are applications for both large and narrow field-of-view telescopes in electromagnetic astronomy.
We analyze the prospects for GW detection and characterization using our clock based scheme, including a comparison of the sensitivity of this technique to other proposed space-based detectors. We highlight new and complementary GW measurement capabilities provided by space-based optical atomic clocks, and discuss the prospects for integrating our scheme with existing proposals.

While there is little doubt that LIGO and other terrestrial detectors will observe numerous additional GW events in the coming years, terrestrial detectors are only sensitive to GWs with frequencies above $\sim$10 Hz, due to seismic and Newtonian noise \cite{LIGOdetection,LIGO,LISA-older,LISA}. The desire to observe a wider range of astrophysical phenomena over longer length and time scales has motivated proposals of larger scale, space-based GW detectors \cite{LISA-older, LISA, AtomInterGW, AtomInterGW2,Yu2011,Yu2015,LoebGWclocks,Vutha}. There are a wide variety of existing and proposed techniques \cite{LIGOdetection,LIGOdetection2,LIGO,TorsionBar,SeismoGW,detweiler1979pulsar,GPS,Ulysses,Apollo,DopplerReview,TerrestrialAIGW}, all of which rely on the same GW effect, namely the periodic change in proper distance between two points in space \cite{Congedo2013}. This effect results in modulation of the arrival times of photons sent over an electromagnetic baseline, which corresponds to effective changes in relative position and velocity. The differences between the various techniques lie in the detection methods, the physical quantity that is being locally measured, and the susceptibility to different noise sources, making particular schemes better suited for specific GW frequency ranges.

The existing and proposed space-based GW detectors can be broadly classified in two categories. The first are optical interferometric detectors analogous to LIGO in space, such as the proposed Laser Interferometer Space Antenna (LISA) \cite{LISA-older} and Evolved-LISA (eLISA) \cite{LISA}, which would be composed of three spacecraft forming either a two or three arm Michelson interferometer, with roughly equal length arms to reduce susceptibility to laser frequency noise. These GW detectors rely on large photon fluxes to split the optical interference fringe down to the required sensitivities, and detect signals in a broad frequency band determined by the detector arm length and residual acceleration noise of the satellites \cite{LISA-older, LISA}.

\begin{figure*}
\begin{center}
\includegraphics[width=1\textwidth]{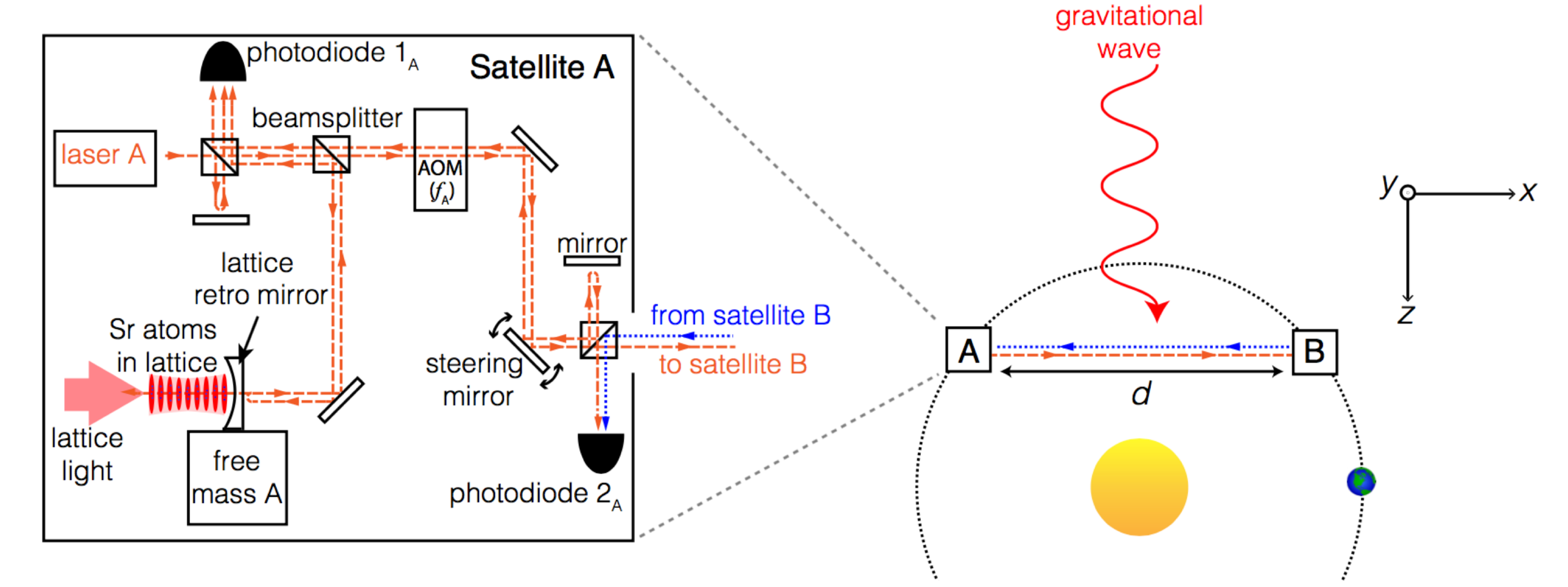}
\caption{ \textbf{Proposed gravitational wave detector (not to scale).} Our detector consists of two identical drag-free satellites, A and B, separated from each other by a distance $d$ along the $x$-axis. Each satellite contains a free-floating reference mass, an ultra-stable laser, and a strontium optical lattice clock. A mirror is mounted on the free mass and is used to define the standing wave of light forming the optical lattice and confining the Sr atoms. Some of the laser light from satellite A (orange, dashed line) is sent to satellite B. The light first passes through an acousto-optic modulator (AOM) driven at frequency $f_{A}$, which offsets the frequency of the light reaching photodiode $2_B$ in satellite B and enables the phase locking of laser B to laser A through heterodyne detection. Vibrations and thermal drifts of the optics on each satellite can be corrected locally by feeding back on the beat notes at $2f_{A,B}$ on photodiodes $1_{A,B}$. Light from laser B (blue, dotted line) is sent back to satellite A to verify the phase lock, to maintain pointing stability, and to enable operation in the reverse mode, with laser A locked to laser B. A plus-polarized gravitational wave propagating along the $z$-axis induces relative motion between the two free masses (see Appendix \ref{App:Doppler}), which can be detected using a clock comparison measurement protocol. The satellite configuration and orbit shown here is intended only for illustration of the basic concepts of our detector. A more sophisticated orbital pattern could be employed to increase the rate of rotation and sweep the detector pattern over a larger region. Additional satellites and optical links could also be used for improved sensitivity and localization of GW sources.}
\label{fig:DetectorDiagram}
\end{center}
\end{figure*}

The second class of space-based GW detectors rely on stable internal frequency references, such as Doppler tracking of distant spacecraft \cite{GPS,Ulysses,Apollo,DopplerReview}. These detectors search for changes in the frequency of electromagnetic waves due to effective Doppler shifts arising from passing GWs. Doppler tracking of spacecraft has been successfully employed to set the existing limits on milliHertz gravitational wave events \cite{GPS,Ulysses,Apollo,DopplerReview}. Because the sensitivities of this class of detector are generally limited by the stability of the frequency reference rather than the photon flux \cite{DopplerReview,CordesShannon}, there is a clear motivation to improve the internal frequency references used for GW detection through the adoption of atomic physics techniques, such as either atomic interferometry (AI) \cite{AtomInterGW,AtomInterGW2,Yu2011,Yu2015}, or optical lattice atomic clocks, as described here.

A GW detector composed of two satellites carrying optical lattice atomic clocks and sharing a single laser over an optical link can measure shifts in the rate of optical phase change by comparing the laser frequency to an atomic degree of freedom on both ends of the baseline. Such a detector is therefore similar to the Doppler tracking method of GW detection \cite{GPS,Ulysses,Apollo,DopplerReview,Vutha}, in that it is sensitive to changes in the apparent relative velocities of the reference masses, rather than changes in the apparent relative distance. An important advantage of the atomic clock scheme over other Doppler tracking methods is that one has full control over the frequency references. As a result, synchronized measurement sequences can be applied at both ends of the baseline to cancel laser frequency noise \cite{CorrNoiseWineland,CorrNoiseKatori,Takano,CorrNoiseTravis,AtomInterGW2}: thus the atomic clock technique requires only two spacecraft, not three, and the differential measurement is entirely limited by the internal atomic transition, not the stability of the local oscillator used to probe it \cite{ClockSr,ClockYb,Katori2015,ClockBetter,PTBion,DopplerReview}. Furthermore, dynamical decoupling (DD) control sequences can be applied to the internal states of the atoms \cite{Bishof,DD1}, extending the range of GW frequencies to which the detector can be maximally sensitive, from milliHertz to tens of Hertz, without requiring any physical changes to the detector. This key feature provides a tunable, narrowband GW detector for tracking evolving GW sources such as inspiraling black hole or neutron star binaries, and bridging the spectral gap between space-borne and terrestrial optical interferometric GW detectors \cite{LIGOdetection,LIGOdetection2,LIGO,LISA-older,LISA}.

Due to the Doppler-based measurement scheme and high quality factor of atomic clock transitions, the optical power requirements on the link between satellites differ from those of optical interferometer GW detectors, as discussed below. In addition, quantum techniques such as atomic spin-squeezing and entangled states \cite{SpinSqueeze,BestSqueezing,ClockGHZ} offer the potential for future improvements in sensitivity, detection bandwidth, and spectral range. Finally, because optical atomic clocks are currently the most accurate frequency references \cite{ClockSr, ClockBetter, ClockYb, Katori2015,PTBion}, and can provide improved sensitivity to beyond-Standard-Model phenomena that may couple to atomic properties such as mass, charge, and spin \cite{DarkMatterClocks0,DarkMatterClocks2,arvanitakiGWDM,IsotopeShifts}, there is already considerable motivation to develop space-hardy optical clocks, and to integrate them with other proposed GW detectors.

\section{Sensing gravitational waves using optical lattice atomic clocks}
Our proposed GW detector, illustrated in Fig.~\ref{fig:DetectorDiagram}, consists of two drag-free satellites in heliocentric orbit (A and B), separated by a length $d$ and connected over a single optical link using conventional optical telescopes. Each satellite contains its own optical lattice atomic clock \cite{Katori2015,ClockYb,ClockSr,SpaceClock}, and its own ultra-stable laser \cite{Laser}. The laser in satellite B is kept phase locked to the light sent from satellite A over the optical link, such that the two lasers function as a single ultra-stable clock laser shared between the two satellites. In each satellite the lattice confining the clock atoms is created using the standing wave formed by retro-reflecting a magic wavelength laser \cite{Katori2015,ClockYb,ClockSr} off of a mirror mounted on a free-floating reference mass, such that the atoms are strongly confined in the reference frame of the free mass and are therefore in free-fall, despite their confinement. Drag-free masses have been studied in great detail by the LISA collaboration, and this technology is currently undergoing testing and verification in the LISA Pathfinder space mission \cite{Pathfinder,Pathfinder2}. The phase of the clock lasers in each satellite is kept referenced to the same mirror using interferometry \cite{FiberPhaseNoise} to cancel out any relative motion of the lasers or optics with respect to the atoms. To cancel the radiation pressure exerted on the free mass by the lattice and clock beams, a set of equal power, counter-propagating lasers are incident on the opposite sides of the free masses. For a 1 kg mass and a 1 W lattice beam, the remaining acceleration noise from the quantum radiation pressure shot noise of the lattice, clock, and compensation beams is far below the GW detector noise floor at frequencies of interest \cite{RPNLISA}.

Operation of the GW detector consists of a synchronous comparison between the two optical lattice atomic clocks. The frequency of laser A is compared to the clock transition in the atoms in satellite A using spectroscopic read-out, such as Ramsey spectroscopy. Synchronization signals are transmitted to satellite B, so that an identical measurement is performed on the atoms in satellite B using laser B, which is phase locked to laser A. Both Ramsey measurements are performed with the same interrogation time $T$. The Ramsey phases accumulated by the atoms in each satellite are recorded, and can then be compared over a standard communication channel. Because the satellites effectively share a single laser and the two measurements are offset by the time required for the laser light to travel from A to B, any laser frequency noise will be common mode for the two measurements, resulting in the same additional acquired phase in each clock, and will thus be rejected. This method of laser frequency noise rejection has been previously utilized in optical atomic clocks to cancel laser noise arising from the Dick effect, and thereby achieve the quantum projection noise limit \cite{CorrNoiseWineland,CorrNoiseKatori,CorrNoiseTravis}; it has also recently been proposed for use in AI-based GW detectors \cite{AtomInterGW2}.

A passing plus-polarized GW of strain amplitude $h$ and frequency $f_{\text{GW}}$, propagating along the $z$-axis  perpendicular to the optical link between the satellites, will periodically change the apparent distance between the free masses A and B, as measured by the null geodesic of the optical link. If the light sent from satellite A to B is used as reference clock light, its frequency will experience a Doppler shift that indicates the GW induced effective relative motion of the two satellites. Hence the atoms in satellite B will experience a local oscillator of a different optical frequency than the atoms in satellite A, and will accumulate a different Ramsey phase. When the two clocks are compared, they will appear to have ``ticked'' at different rates, with the maximum fractional frequency difference between the two clocks given by
\begin{equation}
s \equiv \frac{ \delta \nu}{\nu} =  h \left | \sin \left( \pi f_{\text{GW}} \frac{d}{c}  \right) \right |,
\label{freqdistdep}
\end{equation}
\noindent where $c$ is the speed of light (see Appendix \ref{App:Doppler} for derivation). Note that $s=h$ for the optimal clock spacing $d = \lambda_{\text{GW}}/2$, where $\lambda_{\text{GW}}=c/f_{\text{GW}}$ is the GW wavelength. When compared to an optical interferometric GW detector such as LISA \cite{LISA,LISA-older}, with total optimized arm length $d$, the fractional frequency difference $s$ between two optimally spaced clocks will be equivalent to the fractional change in differential arm length experienced by the optical interferometer. At GW frequencies other than the optimal frequency the magnitude of the detectable signal is determined by the inherent sensitivity of the specific setup, as captured by the detector's transfer function ${\cal T}(f)$ \cite{DopplerReview,TransferFunction} and susceptibility to noise (see Appendix \ref{App:Sensitivity}). As discussed below, the noise floor of optical interferometric detectors is fundamentally limited by white phase noise arising from photon shot noise \cite{LISA}, while the noise floor of the clock detector is dominated by white frequency noise arising from atom projection noise \cite{ClockReview2}. This fundamental physical difference motivates the present consideration of the former as a detector of changes in phase, and the latter as a detector of changes in frequency, so that the detector transfer functions can be directly compared. We emphasize that while both types of detector can in principle express their measurement in terms of either phase or frequency, the respective fundamental physical noise floors and signal to noise ratios will be unchanged.

The transfer function ${\cal T}_{\phi}(f)$ for optical interferometric GW detectors such as LISA is frequency independent for GW frequencies below $c/2d$, but scales as ${\cal T}_{\phi}(f)\propto1/f_{\text{GW}}^2$ at higher frequencies where the photon transit time is longer than a half period of the GW \cite{LISA-older,LISA}. Because the 1~mHz - 1~Hz frequency range is of primary interest for space-based detectors \cite{LISA-older,LISA,AtomInterGW,AtomInterGW2}, ${\cal T}_{\phi}(f)$ sets a maximum arm length for an optical interferometer on the order of $\sim1\times10^9$ meters. In contrast, because the clock GW detector compares the local laser frequency at the two satellites and is thus only sensitive to the effective relative velocity of the satellites, the transfer function ${\cal T}_{\nu}(f)$ of the clock GW detector scales as ${\cal T}_{\nu}(f)\propto f_{\text{GW}}^2\times{\cal T}_{\phi}(f)$ due to the time derivative relating position (phase) to velocity (frequency). Therefore, ${\cal T}_{\nu}(f)\propto f_{\text{GW}}^2$ for $f_{\text{GW}}<c/2d$, but is frequency independent at higher frequencies\footnote{Here we have set aside the ``blind-spot'' frequencies, present for all optical GW detecters, that occur when $\lambda_{\text{GW}}=d$.}\cite{DopplerReview}. We are interested in GW frequencies of $\sim$mHz and above, thus we propose a clock GW detector with a baseline length $d = 5\times10^{10}$ m, setting the minimum frequency that can be detected at the detector's peak sensitivity to be $c/2d\approx3$~mHz. Note that a LISA-like baseline length of $5\times10^9$ meters could be used for the  clock GW detector without sacrificing sensitivity at GW frequencies above $\sim$30~mHz.

\section{Expected sensitivity}
The optical lattice clock GW detector is fundamentally limited by quantum projection noise of the atomic read out, which determines the stability of the differential frequency measurement. We consider two clocks separated by a distance $d$, sharing a clock laser over the optical link with a laser linewidth $\Delta_L$, which is limited by current optical cavity technology to $\Delta_L\ge20$~mHz, an order of magnitude broader than the natural atomic line width $\Delta_A$ \cite{Laser,Bishof}. We assume that the clocks are atom projection noise limited, and that there is perfect single shot readout of each atom's final internal state following the Ramsey sequence. As the two clocks effectively share a single clock laser, laser noise is common mode and the Ramsey free precession time $T$ can be extended considerably beyond the laser coherence time \cite{CorrNoiseWineland,CorrNoiseKatori,CorrNoiseTravis}. Here we assume $T$ can be pushed out to the radiative lifetime of the clock transition, $T_{max}=1/\left(2\pi\Delta_A\right)$. Note that reaching $T_{max}$ also requires the suppression of atomic interactions to avoid collisional broadening and many-body losses, which can be accomplished by loading the atoms into a 3D optical lattice with one atom per site. In addition we assume that the atom lifetime in the lattice exceeds $T_{max}$. Thus for $N$ atoms in each clock and a series of optimized Ramsey measurements, each with precession time $T_{max}$, and a total measurement time $\tau$, the smallest detectable fractional frequency difference $\sigma_{min}$ between the two clocks, and hence the smallest measurable GW-induced strain with our scheme, is given by

\begin{equation}
\sigma_{min}  \left(\tau \right)=\frac{\delta \nu_{min}}{\nu}\bigg|_{\tau}=\frac{\sqrt{\Delta_A}}{\nu \sqrt{2\pi \tau N}},
\label{clock_sense}
\end{equation}

\noindent where $\nu$ is the frequency of the optical clock transition \cite{ClockReview2}. To analyze the achievable GW sensitivity using this technique, we consider a next-generation strontium-87 optical lattice clock, as $^{87}$Sr has the narrowest demonstrated clock transition linewidth~\cite{ClockSr,ClockReview2}. The $^{87}$Sr $^{1}S_0-{}^{3}P_0$ clock transition is at $\nu=430$ THz, and the transition linewidth is $\Delta_A=1$~mHz, yielding $T_{max} = 160$~s~\cite{BoydSr}. Current work is experimenting with the loading of $10^4 - 10^5$ $^{87}$Sr atoms from a degenerate Fermi gas into a 3D optical lattice~\cite{Campbell} to achieve record-long coherence times. With improved lattice power and engineering, one may expect a strontium optical lattice clock to operate with $\sim1\times10^{7}$ atoms. Taking $N=7\times10^{6}$ atoms yields a minimum detectable fractional frequency difference of $\sigma_{min}=1.1\times10^{-20}/\sqrt{\mathrm{Hz}}$. Although this represents a 4 order of magnitude improvement over demonstrated clock stability~\cite{ClockBetter}, the use of correlated noise spectroscopy, along with anticipated large improvements in the atom number, coherence time, and improved laser linewidth, will help realize this gain. Note that because $\sigma_{min}$ can only be achieved using measurements with optimal Ramsey precession time $T_{max}=160$~s, our detector is spectrally narrowband and thus is not well suited for the detection of short burst GWs. Nonetheless, as we discuss below, our detector should be well suited for the detection of GWs emanating from a variety of continuous, spectrally narrow sources, such as compact binary inspirals.

\begin{figure*}
\begin{center}
\includegraphics[width=1\textwidth]{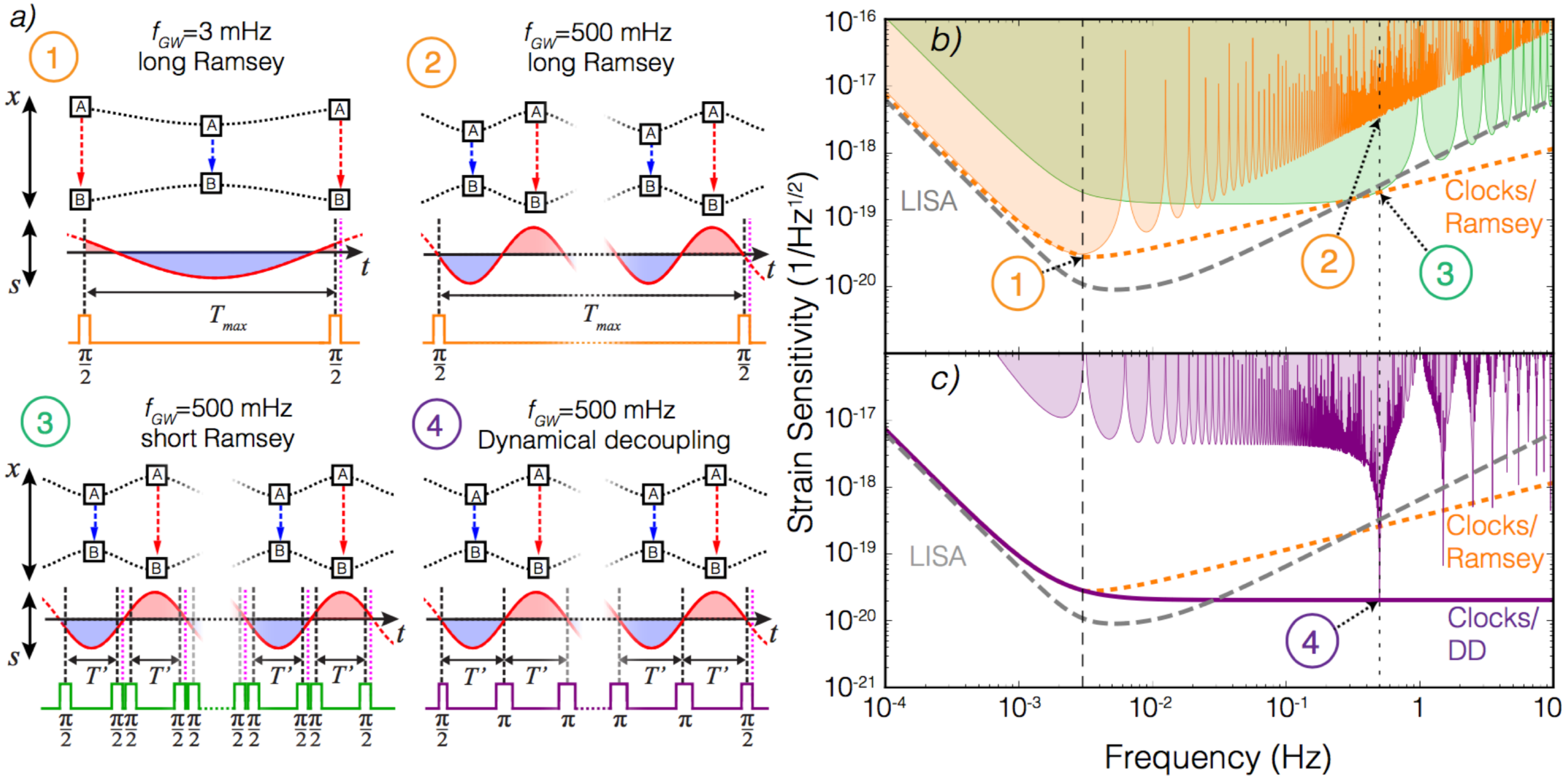}
\caption{\textbf{Measurement protocols and comparison of gravitational wave sensitivities.} \textbf{\textit{a)}} \textit{(1)} A Ramsey pulse sequence performed on the atomic clock transition can be used to detect a GW with $f_{\text{GW}}=3$~mHz, for which a half period matches the atomic linewidth limited interrogation time $T_{max}=160$ s. From top to bottom, we depict the effect of a GW on the relative position of the two satellites (A and B), the sign of the Doppler shift induced on the transmitted laser light, the accumulated clock signal $s$ (see Eq. 1), and the  pulse sequence (pink dotted line indicates atom state readout). \textit{(2)} The same Ramsey pulse sequence as in panel 1 will measure a reduced signal for a GW of frequency $f_{\text{GW}}=500$~mHz, because the fast Doppler shift oscillations will average out. \textit{(3)} A series of shorter Ramsey sequences with $T'=1$ s can be used to detect a $f_{\text{GW}}=500$~mHz GW, with a reduced sensitivity due to the shorter coherent interrogation time. \textit{(4)} A DD sequence with 159 periodic $\pi$-pulses separated by $T'=1$~s can instead be employed to detect a GW with $f_{\text{GW}}=500$~mHz, resulting in the same total accumulated signal $s$ as the Ramsey measurement for $f_{\text{GW}}=3$~mHz as shown in panel 1. \textbf{\textit{b)}} Noise-limited strain sensitivity of our detector to a monochromatic GW using a Ramsey sequence with interrogation time $T_{max}=160$ s (orange filled region), and a Ramsey sequence with $T'=1$ s (green filled region). The Ramsey sensitivity envelope for optimized Ramsey sequences at each GW frequency is shown (orange dashed line), and the projected strain sensitivity of LISA is plotted for comparison \cite{LISA-older}. The clock GW detector consists of one clock per satellite, each with $7\times10^6$ atoms, with the baseline length optimized for $f_{\text{GW}}\ge3$~mHz, giving $d=5\times10^{10}$ m. \textbf{\textit{c)}} Noise-limited strain sensitivity of our detector to a monochromatic GW using a DD sequence with 159 periodic $\pi$-pulses with total interrogation time $T_{max}=160$ s (purple filled region), and the DD sensitivity envelope for optimized DD sequences at each GW frequency (thick purple line). In both Fig.~2b and 2c the strain sensitivities corresponding to panels 1-4 in Fig.~2a are highlighted. The clock GW detector has narrow regions of reduced sensitivity for each measurement sequence when the time between pulses, ($T_{max},T'$), is an integer multiple of a GW period. The polarization and direction of propagation of a GW can also change the measurable signal for LISA and the clock GW detector. All sensitivity curves and envelopes are averaged over all polarizations and directions of propagation of the GWs (see Appendix \ref{App:Sensitivity}).}
\label{fig:Sensitivities}
\end{center}
\end{figure*}

\section{Detector noise floor and dynamical decoupling sequences}
At low frequencies the detector noise floor will be dominated by residual acceleration noise of the free reference masses. This noise has been carefully analyzed by the LISA collaboration \cite{LISA,LISA-older,Pathfinder,Pathfinder2}, and a $1/f^2$ scaling of the sensitivity is anticipated up to a frequency cutoff of $\sim$3 mHz. Our detector experiences the same acceleration noise, resulting in the same scaling of the signal-to-noise ratio. At higher frequencies atom projection noise dominates, which for fixed $T$ is frequency independent. However, optimal sensitivity, $s=\sigma_{min}$, is achievable using a Ramsey sequence only if half the period of the GW is longer than or equal to $T_{max}$, as shown in panel 1 of Fig.~\ref{fig:Sensitivities}a. As illustrated in panel 2 of Fig.~\ref{fig:Sensitivities}a, the signal from higher frequency GWs will partially average out over the course of a single Ramsey measurement of length $T_{max}$, giving rise to reduced sensitivity at higher frequencies. For this particular spectroscopic read out, the reduction scales as $1/f$, just as for an optical interferometry based detector. However, the spectroscopic sequence can be changed on demand without any additional changes to the detector. In order to search for GWs of frequency $f_{\text{GW}}>1/2T_{max}$, a Ramsey interrogation time of $T'=1/2f_{\text{GW}}$ can be used (panel 3 of Fig.~\ref{fig:Sensitivities}a). This results in a reduction in GW sensitivity for all frequencies $f_{\text{GW}}\ge1/2T_{max}$ by a factor of $\sqrt{T'/T_{max}}$ due to the shorter coherent interrogation time. Therefore the strain sensitivity envelope for optimally chosen Ramsey sequences at each $f_{\text{GW}}$ scales as $\sqrt{f}$ at high frequencies, as shown by the dashed orange line in Fig.~\ref{fig:Sensitivities}b and c.

Fortunately, quantum metrology techniques can be applied to achieve optimal, frequency-independent GW sensitivity at higher frequencies. As illustrated in panel 4 of Fig.~\ref{fig:Sensitivities}a, by using a dynamical decoupling (DD) sequence consisting of a Ramsey sequence combined with a train of periodically spaced $\pi$-pulses matched to the frequency of the GW \cite{Bishof,DD1}, it is possible to remain sensitive to a GW with frequency $f_{\text{GW}}>1/2T_{max}$ while still interrogating for $T_{max}$, such that the sensitivity is still given by Eq.~\ref{clock_sense} in a narrow frequency band around $f_{\text{GW}}$. Assuming high-fidelity $\pi$-pulses can be performed, the optimal GW sensitivity $\sigma_{min}$ can be reached for frequencies up to the Rabi frequency for the clock transition, which for conservative local clock laser intensities can exceed $\Omega_{max}\approx100$ Hz. Utilizing DD sequences with $\sim3\times10^3$ $\pi$-pulses or fewer, it is therefore possible to remain maximally sensitivity to GWs with frequencies up to $\sim10$ Hz. These DD sequences are similar in spirit to the ``signal recycling" cavity that is used in the LIGO GW detector to enhance sensitivity in a tunable narrow bandwidth \cite{LIGO,Recycling}; however such a recycling scheme is impossible for a much longer baseline space-based optical interferometer like LISA due to optical diffraction. We note that similar ``resonant'' pulse sequences have also been recently proposed for use in AI detectors \cite{ResonantAI}. The broad frequency range over which our proposed clock GW detector can be tuned and remain maximally sensitive is well suited for the study of binary inspirals and mergers, which chirp upward in frequency as the two bodies spiral inwards at an increasing rate \cite{LIGOdetection,LIGOdetection2}. Once an on-going GW event has been detected, the spacing of the $\pi$-pulses in the DD detection sequence can be ``chirped'' along with the signal to remain optimally sensitive to the particular event throughout its evolution.

\section{Comparison to other proposals}
In Fig.~\ref{fig:Sensitivities}b and \ref{fig:Sensitivities}c we plot a comparison of the strain sensitivities of the clock GW detector and the proposed LISA mission \cite{LISA-older, LISA}, using Ramsey and DD sequences respectively. The LISA GW detector uses optical interferometry, a proven concept for which extensive design and testing has already been performed. Furthermore, the LISA GW detector will provide broadband sensitivity as plotted in Fig.~\ref{fig:Sensitivities}b, in contrast to our clock GW detector, which makes a narrowband measurement at a frequency selected by the applied control sequence. We therefore consider our proposal as complementary to the LISA mission; we envision that an optical clock GW detector could be integrated with, and operated in parallel to, a LISA optical interferometer without reducing the sensitivity of either GW detector. As an example of the advantages of such a hybrid detector, following the detection of an on-going binary inspiral at mHz frequencies by LISA, the clock GW detector would enable continued observation of the event as the frequency rises out of the detection bandwidth of LISA, all the way through the final moments of the merger, or until it becomes detectable by terrestrial GW detectors. In addition, LISA could greatly benefit if next-generation optical lattice atomic clocks are made ready for space, as the ultra-stable lasers locked to the clocks would provide the best possible local oscillator for the optical interferometer.

Atomic interferometer (AI) GW detectors have been proposed with comparable predicted sensitivities to both LISA and our optical atomic clock proposal, for similar baseline lengths, and requiring only two satellites \cite{AtomInterGW,AtomInterGW2}. Importantly, the atoms in an AI GW detector are completely unconfined, and hence there is no need for drag-free reference masses as the atoms themselves are in free-fall. However, the AI proposal also requires that the atoms be cooled to picoKelvin temperatures \cite{picokelvin}, as the measurement is made using the motional states of the atoms. In contrast, our clock-based scheme requires drag-free satellite technology, but this enables the loading of atoms at microKelvin temperatures into the ground state of the optical lattice \cite{ClockSr}. Furthermore, other than the recent ``resonant'' AI detector \cite{ResonantAI}, to date AI proposals have primarily focused on a measurement scheme that involves repeatedly imprinting the phase of the optical field onto the motional degrees of freedom of the atoms using light propagating back and forth between the satellites, ultimately yielding an anticipated sensitivity curve more similar to that of optical interferometric GW detectors than that of Doppler-shift based detectors \cite{AtomInterGW,BakerThorpe}.

\section{Analysis of optical power requirements}
Photon shot noise is a considerable fraction of the noise budget of other space-based GW detector proposals \cite{LISA}. We now analyze the requirements on transmitted optical power so that the sensitivity of the optical atomic clock GW detector is limited only by atom projection noise at frequencies above 3 mHz. We restrict our analysis to the fundamental case of GW detection using Ramsey sequences. Photon shot noise enters the clock GW detector through noise on the phase-locked loop (PLL) used to lock laser B to the light arriving from laser A, with the phase error variance of the loop given by

\begin{equation}\label{PLL_noise}
\delta\phi^2 \approx \frac{h\nu B}{\eta P_B} + \frac{\Delta_L}{B},
\end{equation}

\noindent where $P_B$ is the power from laser A that is received at satellite B, $\eta$ is the detector quantum efficiency, $\Delta_L$ is the linewidth of each laser, and $B$ is the PLL bandwidth \cite{PLL}. The first term results from photon shot noise on the optical link, while the second term arises from phase excursions of laser B due to the finite loop bandwidth. To keep lasers A and B coherent at all times the PLL must not undergo phase cycle slips, requiring $\delta\phi^2\ll1$. An additional requirement is that the optimal loop bandwidth  $B_{opt} = \sqrt{\eta P_B \Delta_L/(h \nu)}$, found by minimizing $\delta\phi^2$, must be larger than the GW frequency to be detected so that the loop can respond to the GW signal. In the limits of long measurement time and large Rabi frequency relative to the optimal loop bandwidth, $B_{opt}\tau\gg1$, $B_{opt}\ll\Omega_R$, the noise floor due to both atom projection noise and photon shot noise for a continuous series of uninterrupted Ramsey measurements is then given by (see Appendix \ref{App:Power} for derivation)

\begin{equation}\label{combined_noise}
\sigma^2(\tau) = \frac{1}{(2 \pi \nu)^2 T \tau} \left( \frac{1}{N} + \frac{T}{ \tau} \sqrt{\frac{h \nu \Delta_L}{\eta P}} \right).
\end{equation}

Because the photon shot noise (second term) in Eq.~\ref{combined_noise} scales with $1/\tau^2$, a well known result for Doppler tracking GW searches \cite{DopplerReview}, while the atom projection noise term scales as $1/\tau$, the atom projection noise will dominate over photon shot noise at sufficiently long averaging times, and Eq.~\ref{combined_noise} will reduce to Eq.~\ref{clock_sense} when $T=T_{max}$. For example, taking $\Delta_L=30$~mHz, $\eta=0.5$, and the averaging time to be at most 1~day, we find that the received optical power at satellite B must exceed $P_B\gtrsim3$~pW in order for the clock detector to be atom projection noise limited. At 1 day of averaging, the minimum detectable strain of a continuous GW with a frequency between 3 mHz and 10 Hz would then be $h_{min}\approx3.7\times10^{-23}$. For long optical baselines, the power received at satellite B is related to the power transmitted from satellite A by $P_B=P_A\left(\pi R^2 \nu/d c\right)^{2},$ where $R$ is the radius of the telescope used on both satellites \cite{TransferFunction}. Hence for $R=30$ cm and the proposed satellite separation of $d=5\times10^{10}$ m, the clock GW detector requires a transmitted power of $P_A \gtrsim 50$~mW. If the clock GW detector sensitivity were to be improved by increasing the atom number, the full gain in sensitivity could be realized by either increasing the optical power in order to reach the lower projection noise floor in the same averaging time, or by simply averaging for longer.

The noise floor given in Eq.~\ref{combined_noise} is for a series of continuous Ramsey measurements with no dead time, which could be achieved through the interleaved operation of two clocks on each satellite \cite{KasevichTwoClocks,LudlowTwoClocks,Meunier}. However, if we restrict the detector to a single optical lattice clock per satellite, detector operation may require a small but finite dead time between subsequent measurements, which can introduce additional susceptibility to differential laser noise through a process known as the Dick effect \cite{DickEffect}. Because the PLL is kept running continuously, it can bridge the dead time between subsequent Ramsey sequences, suppressing the differential laser noise so long as the loop bandwidth is kept above the Rabi frequency, which acts as a lowpass filter for the atomic response. However, this places additional requirements on the optical power received at satellite B. In particular, if $\Omega_{R}\ll B_{opt}$, and in the limit of sufficiently large dead time $T_D\gg1/B_{opt}$ (see Appendix \ref{App:Power}), Eq.~\ref{combined_noise} becomes

\begin{equation}\label{eq:Dick}
\sigma^2 = \frac{1}{(2 \pi \nu)^2 T \tau } \left( \frac{1}{N} + \frac{2}{r}  \frac{h \nu}{ \eta P_B}\Omega_{R} \right),
\end{equation}

\noindent where $r=T/(T+T_D)$ is the duty cycle. This yields the intuitive condition that in order to remain atom projection noise limited the number of photons received at satellite B during the Ramsey control pulses must be larger than the number of atoms $N$ used in each run of the measurement, bounded by the condition for high fidelity $\pi/2$ pulses, $\Omega_{R}\gg\Delta_L$. Taking $\Delta_L=30$~mHz, $\Omega_R=1$~Hz, $\eta=0.5$, and $r=0.9$, we find that the received optical power at satellite B must exceed $P_B\gtrsim10$~pW, which for the detector dimensions given previously yields $P_A\gtrsim150$~mW, comparable to the $\sim$1~W of transmitted power required by the LISA detector \cite{LISA-older,LISA}.

\section{Sources of future improvement}
While our proposal already offers competitive sensitivities in a complementary frequency range to other proposed space-based GW detectors, there are also potential upgrades that can be anticipated to further improve detector performance. For example, while only two satellites and a single optical baseline are fundamentally necessary to make our detector operational, there are a number of scientific advantages to using more arms or an array of two-arm detectors. A clock network composed of a distributed array of spacecraft with phase coherent optical links between nearest neighbors could enable clocks in space to be compared over considerably longer distances than a single baseline scheme, and could also provide optimal sensitivity for arbitrarily polarized GWs propagating in any direction, as well as the ability to localize the GW source direction. In addition, as ground-based optical atomic clocks become increasingly precise there is growing motivation to build space-based clocks for metrology, in order to avoid the gravitational redshifts caused by seismic activity \cite{ClockGR,ClockReview2}. We emphasize that our GW detection scheme is compatible with a space-based clock network designed primarily for time-keeping and navigation.

We can also anticipate increases in the detection bandwidth without sacrificing sensitivity by using spin-squeezed and GHZ atomic states \cite{SpinSqueeze,BestSqueezing,ClockGHZ}. These entangled quantum states can be used to bypass the standard quantum limit for short interrogation times, and hence change the sensitivity scaling with atom number from $\sigma_{min} \propto 1/\sqrt{N}$, as given in Eq.~\ref{clock_sense}, to $\sigma_{min}\propto1/N$. To mitigate the photon shot noise restrictions at short averaging times, more sophisticated allocations of the atomic resources using phase estimation protocols could also be employed \cite{ClockGHZ}. Furthermore, in the present proposal we focused exclusively on the 1~mHz linewidth, $^{1}S_0-{}^{3}P_0$ transition in $^{87}$Sr. The use of correlated noise spectroscopy offers the prospect of switching to a different isotope or atomic species with a narrower clock transition, thereby increasing the coherent interrogation time and improving the sensitivity to GWs \cite{Rosenband,ClockNetwork}. Candidate atoms include neutral Mg \cite{MgClock}, or the bosonic isotopes $^{84}$Sr and $^{88}$Sr, where the linewidth of the otherwise forbidden clock transition can potentially be controlled using a second dressing laser \cite{Sr88Dressing}.

\section{Outlook}
We have proposed a gravitational wave (GW) detector consisting of two satellites each containing an optical lattice atomic clock linked by ultra-stable optical laser light over a single baseline. Synchronous clock comparisons will allow detection of GWs via the effective Doppler shift of the shared laser light. With realistic projections for the atomic clock performance, our detector is expected to provide comparable strain sensitivity to that of other proposed space-based GW detectors based on optical and atomic interferometers \cite{LISA-older,LISA,AtomInterGW,AtomInterGW2}, along with several complementary features. In particular, our detector bridges the detection gap between space-based and terrestrial optical interferometric GW detectors through tunable, narrowband GW detection with constant sensitivity over a broad frequency range from $\sim$3 mHz to 10 Hz, while also offering flexible laser power requirements for the spacecraft link, and requiring only readily realizable atomic technology. We therefore anticipate that optical clock GW detectors can play a complementary role to optical interferometer detectors in both first and future generation space-based GW missions. Beyond GW detection, clocks also offer sensitivity to other fundamental physical and astronomical phenomena that may couple to atomic properties such as mass, charge, and spin, including searches for dark matter, violations of fundamental symmetries, and variations of fundamental constants \cite{DarkMatterClocks0,DarkMatterClocks2,arvanitakiGWDM,IsotopeShifts}. Key challenges to be addressed in future works include: the development of optimized clock measurement protocols tailored for GW sources of interest, as well as spectral characterization and detection feasibility studies of known GW sources; the design of space-hardy, high-precision atomic clocks and ultra-stable lasers \cite{SpaceClock}, which will also directly benefit other proposed space-based GW detectors; detailed analysis of  the noise susceptibility of DD sequences requiring many operations \cite{ThompsonWPN}; and the demonstration of quantum metrology techniques involving entanglement to enhance both the sensitivity and detection bandwidth of clock GW detectors.

\begin{acknowledgments}
We thank Avi Loeb and Dan Maoz for providing the initial inspiration to pursue this work. We also thank Johannes Borregaard, Akihisa Goban, Jason Hogan, Mark Kasevich, Edward Marti, Holger M{\"u}ller, Matthew Norcia, Dan Stamper-Kurn, and James K. Thompson for helpful discussions and insights. This work was supported in part by NIST, NASA, JILA PFC, NSF, CUA, NSSEFF, and DARPA. I.P. thanks the NSF for support through a grant to ITAMP. S.K. thanks the NRC postdoctoral fellowship program for support.
\end{acknowledgments}

\appendix
\section{Derivation of the effective Doppler shift induced by a passing gravitational wave}\label{App:Doppler}
A passing gravitational wave (GW) induces periodic changes in the light travel time between emitter and detector\footnote{One can expect an additional effect due to the time dilation induced by the GW itself. However, this effect would be of second order in GW strain amplitude $h$, and is therefore vanishingly small when compared to the sensitivity of current clocks.}. In this section we derive the magnitude of this effect as a function of the GW amplitude, the orientation between the satellites and the direction of propagation of the GW, and the distance between the clocks. Similar analyses have been performed for proposed detectors that utilize Doppler tracking \cite{estabrook1975response} and pulsar timing \cite{detweiler1979pulsar}. Our detection scheme involves only a one-way link as in the case of pulsar timing, but with full experimental control on both sites for the emission and detection of the signal.

\begin{figure}[b]
\centering
\includegraphics[width=0.8\columnwidth]{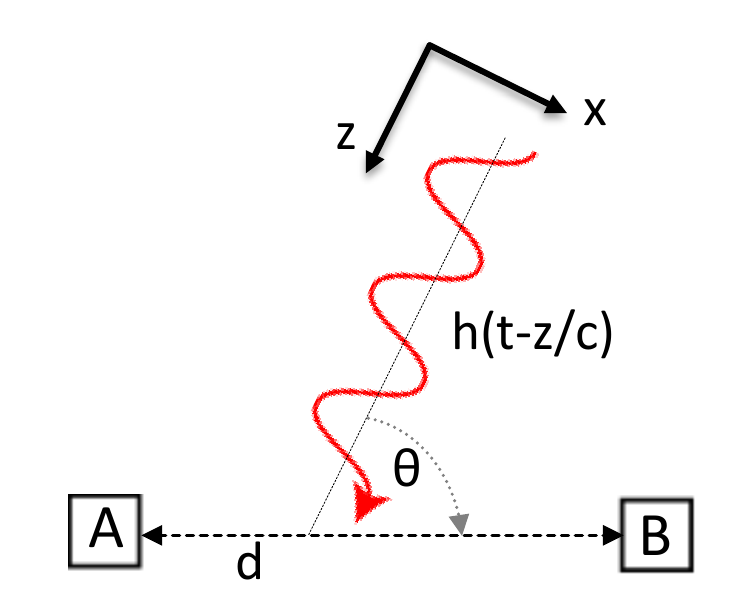}
\caption{\small A GW incident along the $z$-axis periodically changes the light travel distance between $A$ and $B$.}
\label{GRwaves}
\end{figure}
Weak gravitational fields are captured by a perturbed metric $g_{\mu \nu} = \eta_{\mu \nu} + h_{\mu \nu}$, where $\eta_{\mu \nu}$ is the Minkowski metric and $|h_{\mu \nu}| \ll 1$ is a small perturbation. GWs are described in the transverse traceless gauge by the metric:
\begin{equation}\label{eq:metric}
g_{\mu \nu} = \left( \begin{tabular}{cccc}
	$-1$ & 0 & 0 & 0  \\
	0 & $1 + h_+$ & $h_\times$ & 0  \\
    0 & $h_\times$ & $1 - h_+$ & 0  \\
    0 & 0 & 0 & $1$
	\end{tabular} \right),
\end{equation}
\noindent where $h_+(t-z/c)$ and $h_\times(t-z/c)$ correspond to the two polarizations of the wave, which travels in the $z$-direction. For simplicity we first calculate the effect for a plus-polarized plane wave with $h=h_+ =|h| e^{-i 2 \pi f(t-z/c)}$, where $f$ is the frequency of the wave and $|h|$ its amplitude (arbitrary polarizations are restored with the substitution $|h| \rightarrow |h_+| \cos(2 \psi) + |h_{\times}| \sin(2\psi)$, where $\psi$ is the polarization angle). The line element for this metric is then
\begin{equation}\label{eq:ds}
ds^2 = - c^2dt^2 + (1+ h )dx^2  +  (1- h )dy^2  + dz^2.
\end{equation}
We now consider the situation depicted in Fig.~\ref{GRwaves}, where a light signal is sent at time $t$ from system $A$ to system $B$, which is at a distance $d$ in the $x-z$-plane. A light-like curve is defined by $ds^2=0$. Parameterizing the curve by $r$ with $x=r\sin\theta$, $y=0$ and $z=r\cos\theta$, the coordinates for the curve become (to lowest order in $h$):
\begin{equation}\label{eq:dt}
c dt = \left(1 + \frac{1}{2} h \sin^2\theta\right) dr .
\end{equation}
As the signal is emitted at coordinate time $t$ and travels from $A$ to $B$ in a time $t_1=t + d/c$ to lowest order in $h$, it travels an apparent distance
\begin{equation}
D_{AB} = c \int_{t}^{t_1} dt' = \int_{0}^{d}\left(1 + \frac{1}{2} h (1-\cos^2\theta)\right) dr ,
\end{equation}
\noindent where the GW is parameterized by $h=h(t+r/c - r \cos\theta/c )$. In terms of the indefinite integral of the wave, H(t), the above expression becomes
\begin{equation}\label{eq:D01}
\begin{split}
D_{AB} & = c (t_1 - t) = \\ 
& \! \! d + \frac{c}{2} \left( 1+ \cos\theta\right)\left[H(t) - H\left(t+\frac{d}{c}(1-\cos\theta)\right)\right] .
\end{split}
\end{equation}
In flat space the distance traveled by the light would just be given by $d$, but the presence of the GW periodically changes the apparent length of the light path. In Doppler tracking techniques, the signal is reflected back to $A$ and measured there. Here, instead, we consider measurement directly on $B$.
The rate of change gives a Doppler shift of the signal $\sigma \equiv \dot{D}_{AB}/c = \Delta \nu/\nu$, where $\nu$ is the optical frequency:
\begin{equation}\label{eq:Doppler}
\begin{split}
 s = \frac{\Delta \nu}{\nu} & = \frac{1+\cos\theta}{2}  \left[ h(t) - h\left(t+\frac{d}{c}(1-\cos\theta) \right)\right].
\end{split}
\end{equation}
This apparent Doppler shift is the signal to be detected. The effect is maximized for $\theta=\pi/2$, i.e. for the detector aligned perpendicularly to the GW, while the signal disappears for $\theta=0$, i.e. in the direction of propagation of the GW. Similarly to interferometric detection schemes, the frequency shift is due to transversal motion of test bodies as the GW is passing.

From equation \eqref{eq:Doppler}, we can see that when using a single shared local oscillator to compare two clocks positioned a distance $d$ apart in the plane ($\theta=\pi/2$) of a passing GW of amplitude $|h|$ and wavelength $\lambda_{\text{GW}}=c/f$, the clocks will appear to ``tick'' at different rates, with the maximum fractional frequency difference between the two clocks given by
\begin{equation} \label{eq:sensitivity}
s_{max} = |h| \left| \sin \left( \pi \frac{d}{\lambda_{\text{GW}} }  \right) \right|.
\end{equation}
Note that the detector is insensitive to GWs with wavelengths that match a multiple of the baseline $d$.

\section{Detector sensitivity}\label{App:Sensitivity}
\begin{figure}[b]
\centering
\includegraphics[width=0.8\columnwidth]{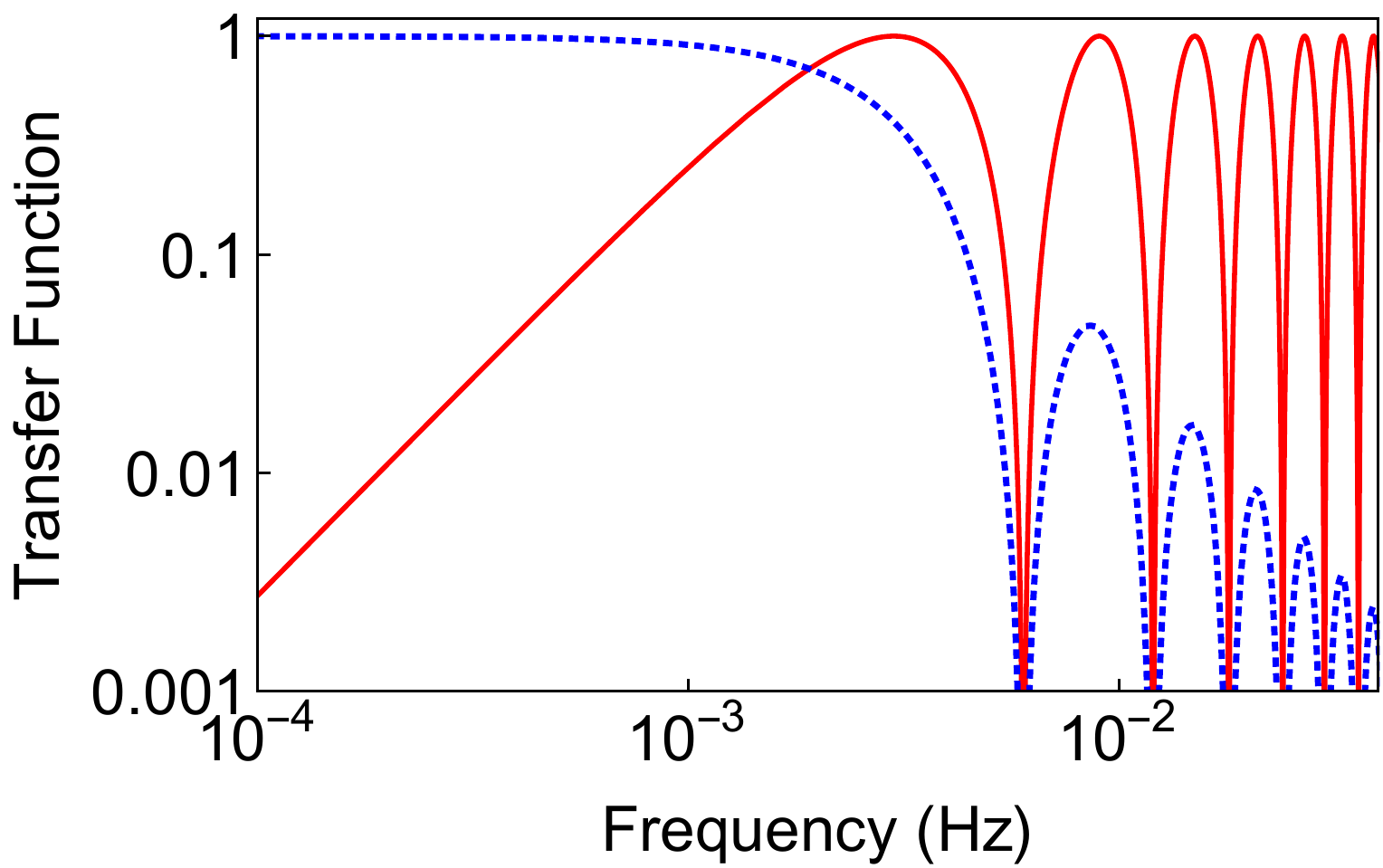}
\caption{\small Transfer functions for a detector sensitive to changes in frequency, ${\cal T}_{\nu}(f)$ (red curve), as compared to a detector sensitive to phase, ${\cal T}_{\phi}(f)$ (blue dotted curve). Frequency measurements yield the maximal signal for $f=(n+1/2)c/d$, $n \in \mathbb{N}_0$, while the sensitivity is drastically reduced for $f < c/(2d)$. In contrast, phase measurements become significantly less sensitive for frequencies $f \gtrsim c/(2d)$, even without the presence of noise. Here the distance between satellites is $d=5\times 10^{10}$~m, as in the main text.}
\label{Fig-Trans}
\end{figure}

For space-based detectors, the effect of geometric factors on the sensitivity is typically described by the transfer function ${\cal T}(f)$, which captures the detector response to specific GW frequencies \cite{cornish2003lisa}. We can express Eq.~\eqref{eq:Doppler} in Fourier-space, which gives
\begin{equation}\label{eq:DopplerFourier}
\tilde{s}(f) = \frac{1}{2} \tilde{h}(f) \left( 1- e^{i 2 \pi f d/c} \right),
\end{equation}
where the tilde denotes the Fourier transform $\tilde{s}(f) = \int dt e^{i 2 \pi f t} s(t)$. The expression multiplying $\tilde{h}$ in Eq.~\eqref{eq:DopplerFourier} depends only on the geometry of the detector and gives rise to its geometric transfer function, which is ${\cal T}_{\nu}(f) = |(1- e^{i 2 \pi f d/c})/2|^2 = \textrm{sin}^2(\pi f d/c)$. It is different than for the case of phase detectors in two ways: we consider only a single one-way link between two satellites, and are sensitive to frequency, i.e. changes in the phase of the light. For detectors sensitive to phase, the additional derivative results in the transfer function ${\cal T}_{\phi}(f) = \textrm{sinc}^2(\pi f d/c)$ \cite{schilling1997angular}. A comparison between the transfer functions of a phase and a frequency detector for an otherwise identical geometry is shown in Fig.~\ref{Fig-Trans}.
%

The actual measured signal for the clock-based  detector depends on the measurement scheme used for the atoms. A long integration time $T$ increases the sensitivity (see Eq.~2 in the main text), but is limited by the atomic linewidth. The signal acquired for a clock measurement between $t_0$ and $t_0+T$ is therefore
\begin{equation}\label{eq:bars}
\bar{s}  =\frac{1}{T} \left| \int_{t_0}^{T+t_0} dt s(t) \right| =\left| \int_{- \infty}^{\infty} dt F(t_0-t) s(t) \right| ,
\end{equation}
\noindent where $F(t)$ is a window function that captures the measurement sequence of duration $T$. For a Ramsey measurement (ignoring the finite pulse durations), the window function is just $F(t) = 1/T$ for $t \in [-T,0] $ and $F(t)=0$ otherwise. For a continuous GW with $h(t)=|h|\sin(2 \pi f t + \varphi)$, this gives
\begin{equation}\label{eq:bars2}
\begin{split}
\bar{s} & = \frac{|h|}{\pi f T} \left| \textrm{sin}\left(\pi f \frac{d}{c}\right) \textrm{sin}(\pi f T) \right. \times \\
&\times  \left. \textrm{cos}\left( \pi f \left(2 t_0+ \frac{d}{c} +T \right)  + \varphi\right) \right|.
\end{split}
\end{equation}
\noindent As we consider a continuous signal, we can adapt the starting time of the measurement to account for $\varphi$ and thus set the argument of the cosine to 0 to give

\begin{equation}\label{eq:bars3}
\bar{s}  = |h| \left|\textrm{sin}\left(\pi f \frac{d}{c}\right) \, \textrm{sinc}(\pi f T)\right|.
\end{equation}

\noindent The sine term in Eq.~\eqref{eq:bars3} captures the light-travel time between the two satellites, while the sinc-function appears due to intergation for a time $T$. Equivalently, we can describe the measurement by a transfer function ${\cal T}_{T}(f) ={\cal T}_{\nu}(f) \textrm{sinc}^2(\pi f T)$, as can also be seen by using directly the Fourier transform of Eq.~\eqref{eq:bars}. Ideal sensitivity is achieved only for the frequency $f=1/(2 T)$ and distance $d=c T$. For higher frequency GWs, the signal strength is reduced due to the finite integration time $T$. Reducing the integration time to $T'<T$ gives an ideal signal at $2 f T' =1$, but causes the atomic clocks to be less sensitive due to atom projection noise.

Using dynamical decoupling allows the detector to be ideally sensitive at frequencies other than $1/(2 T)$. Instead of the integrated signal given in Eq.~\eqref{eq:bars}, the detection is performed with a window function $F_{dd}(t)$, such that
\begin{equation}
\bar{s}_{dd} = \left|\int_{-\infty}^{\infty} dt s(t) F_{dd}(t_0-t)\right|.
\end{equation}
The window function is defined by the particular dynamical decoupling sequence that is utilized. For our purposes, we use the PDD sequence with $n$ $\pi$-pulses, given by $T F_{dd}(-t)=\Theta(t) + 2 \sum_{k=1}^n (-1)^k \Theta(t-k T/n) + (-1)^{n+1}\Theta(t-T)$, where $\Theta(x)$ is the Heaviside step function. Adapting the measurement time such that $\varphi + \pi f d/c = \pi/2$, the signal becomes
\begin{equation}\label{eq:DD}
\begin{split}
\bar{s}_{dd} & = |h| \left|\textrm{sin}\left(\pi f \frac{d}{c}\right) \, \textrm{sinc}\left(\pi f \frac{T}{n}\right) \times  \right. \\
& \qquad \times \left. \frac{1}{n} \sum_{k=1}^{n}\left( -1\right)^{k+1} \textrm{sin} \left( \pi f (2k-1)\frac{T}{n}\right)\right| .
\end{split}
\end{equation}
With DD, the signal is maximized for $f=n/(2T)$, but is reduced for other frequencies. Thus DD is ideal to select a specific frequency at which the detector is maximally sensitive. The minima closest to the main peak occur at $f=(n \pm 1)/(2T)$, we thus define the detector bandwidth as $\Delta f \approx 1/T$. Outside this frequency range the detector can still operate, but with a reduced sensitivity.

Restoring the angular dependence as in Eq.~\eqref{eq:Doppler}, and averaging over all angles and polarizations, we get
\begin{equation}\label{eq:DDav}
\begin{split}
\bar{s}_{dd} & = |\langle h \rangle | \left|\sqrt{\frac{2}{3} - \frac{1}{(2 \pi f \frac{d}{c})^2} + \frac{\textrm{sin}(4\pi f \frac{d}{c})}{2 (2 \pi f \frac{d}{c})^3}} \,  \textrm{sinc}\left(\pi f \frac{T}{n}\right) \times  \right. \\
& \qquad \times \left. \frac{1}{n} \sum_{k=1}^{n}\left( -1\right)^{k+1} \textrm{sin} \left( \pi f (2k-1)\frac{T}{n}\right)\right| ,
\end{split}
\end{equation}
\noindent where $\langle h \rangle  = \sqrt{|h_{\times}|^2 + |h_{+}|^2}/\sqrt{2}$ is the mean GW amplitude.

\section{Derivation of optical power requirements}\label{App:Power}
Our proposed detector utilizes a phase-locked loop (PLL) to lock laser B in satellite B to the light sent from laser A in satellite A, such that the two lasers function as a single ultra-stable clock laser shared between the two satellites. Such a setup allows for correlated noise spectroscopy \cite{CorrNoiseWineland,CorrNoiseKatori,CorrNoiseTravis}, which enables the Ramsey interrogation time $T$ to be extended far beyond the laser coherence time (1 s) out to the atomic radiative lifetime (160 s). While laser frequency noise arising from the laser linewidth $\Delta_L$ can be eliminated using this technique, shot noise on the optical link and the finite bandwidth of the PLL will give rise to relative phase noise between Laser A and B. Here we analyze the power requirements stemming from the individual laser linewidths, dead time between measurements, Rabi frequency, and the shot noise in the PLL. Because of the differential measurement, our system can be viewed as a single clock probed by a laser with noise given by the fractional relative frequency between Laser A and B, $y=\delta \nu / \nu$, and we denote the uncertainty in the relative frequency as

\begin{equation}\label{eq:VarLight}
\sigma^2_{y} = \langle \bar{\delta \nu}^2 \rangle/\nu^2 ,
\end{equation}
\noindent where $\bar{\delta \nu} = (1/\tau) \int_{t_0}^{t_0+\tau} \delta \nu (t) dt$ is the average relative frequency in a measurement window of time $\tau$. The above expression is the true variance of the average frequency, in practice the Allan variance (or two-sample variance) is a more practical measure of the frequency instability \cite{rutman1978characterization}. We can express the integral again as a convolution with a window function $h(t)$, which captures the sensitivity to frequency noise during a measurement of duration $\tau$: $\sigma^2_{y} = \langle \left( \int_{- \infty}^{\infty} dt h(t_0-t) y(t)\right)^2  \rangle$, or in Fourier space
\begin{equation} \label{eq:VarLight2}
\sigma^2_{y} = \int_0^{\infty} df |\tilde{H}(f)|^2 S_{y}(f) ,
\end{equation}
\noindent where we expressed the variance in terms of the one-sided power spectral density (PSD) $S_y(f)$ and the noise transfer function of the measurement, given by the Fourier transform $\tilde{H}(f) = \int dt e^{i 2 \pi f t} h(t)$. The window function $h(t)$ is determined by the applied spectroscopy sequence, including the Rabi frequency and pattern of the applied atomic control pulses, the dead time between subsequent sequences, and the number of averaged measurements. In contrast, the noise spectrum $S_y(f)$ is completely independent of the measurement protocol, and is instead determined by the design of the PLL, the individual laser linewidths, and the optical power received at satellite B.

We first consider $S_y(f),$ using a simple model which captures the main features of a PLL (for a detailed analysis of phase-locked loops and various loop designs, see Refs. \cite{PLL,chen1987comparison}). We assume that the laser phase $\phi^B$ is updated in a time step $t_k$ according to $\phi_{k+1}^B = \phi_{k}^B + \phi_{k}^{corr} $, where $\phi_{k}^{corr}$ is an applied correction based on the outcome of the heterodyne measurement of lasers $A$ and $B$. The demodulated outcome is a signal $i \propto \textrm{sin}(\phi^A-\phi^B)$ with a shot noise contribution $n$. For small phase differences $\delta \phi = \phi^A-\phi^B$ and a loop bandwidth $B$, the correction in the loop is $\phi_{k}^{corr} = B \int^{t_k+1/B}_{t_k}\delta \phi_k + n(t_k)$. Without shot noise, this loop would give $\phi_{k+1}^B \rightarrow \phi_{k}^A$ in the limit of arbitrarily large bandwidth. However, the shot noise restricts the bandwidth, as it increases with larger $B$. For times $t \gg 1/B$, we can treat the steps as infinitesimal and obtain a loop differential equation $\dot{\delta \phi} = - B \delta \phi - \nu + B n(t)$, where $\nu$ is the laser frequency. Writing this in Fourier space, we obtain the noise power spectral density
\begin{equation}\label{eq:PLLPSD}
S_{\varphi}(f) = \frac{\Delta_L}{(2 \pi f)^2 + B^2} + \frac{h \nu B^2}{\eta P_B((2 \pi f)^2 + B^2)} ,
\end{equation}
\noindent where $B$ is the loop bandwidth, $\Delta_L$ is the linewidth of the two lasers, $P_B$ is the received power from satellite A at the PLL photodetector, and $\eta$ is the detection efficiency. The first term is due to white frequency noise from the two laser linewidths, which is suppressed in the PLL within the bandwidth $B$, while the second term is the photon shot noise of the optical link, which sets the noise floor for the heterodyne detection in the PLL.The bandwidth of the loop can be optimized to minimize the additional phase noise, which gives the optimal bandwidth $B_{opt} = \sqrt{\eta P_B \Delta_L/(h \nu)}$.

\begin{figure}[b]
\centering
\includegraphics[width=0.8\columnwidth]{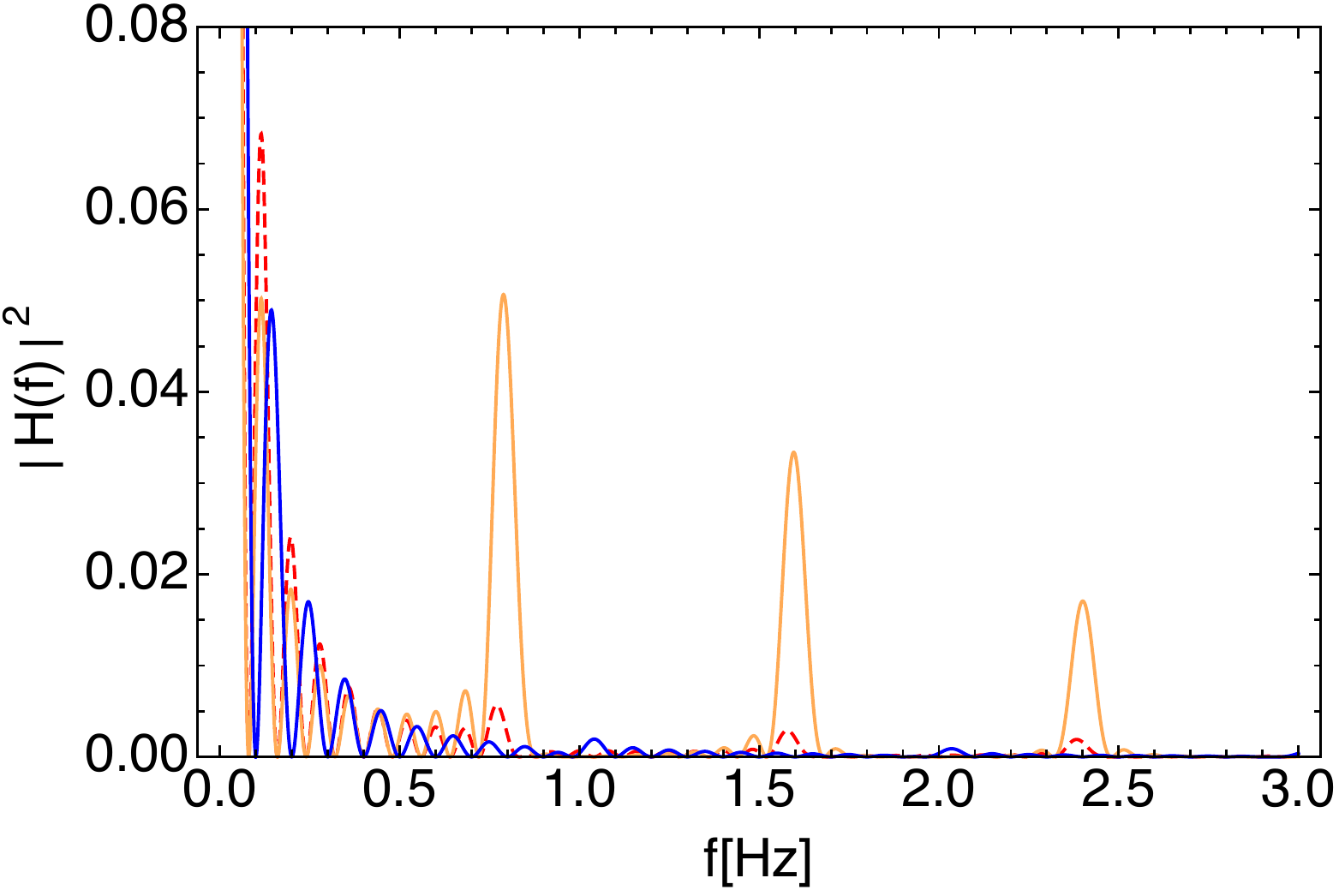}
\caption{\small Transfer function capturing the sensitivity to frequency fluctuations, eq. \eqref{eq:NoiseTrans}. The blue dashed curve shows the case for $n=10$ measurements with no dead time and $\Omega_R = 100$~ Hz. The orange and red curves show the transfer functions for $n=10$, and $r=0.8$, with $\Omega_R = 100$~Hz (orange curve) and $\Omega_R = 10$~Hz (red dashed curve), and the Ramsey time $T=1$~s. Spikes appear due to the Dick effect at frequencies $f=r/T$. The transfer function attenuates frequencies above $\Omega_R$ and thus acts as an effective low-pass filter.}
\label{Fig-TransNoise}
\end{figure}

The noise transfer function $|\tilde{H}(f)|^2$ in Eq.~\eqref{eq:VarLight2} depends on the precise details of the spectroscopy sequence. For the sake of brevity and clarity we restrict our present analysis to Ramsey measurements. We note that for spectroscopic sequences other than Ramsey, additional susceptibility to photon shot noise can be introduced \cite{ThompsonWPN}. DD operation may therefore require additional optical power than Ramsey, and will be studied in detail in future works. The sensitivity function $h(t)$ describes the response of the atoms to frequency fluctuations \cite{santarelli}, and for Ramsey interrogation it is given by $h(t)=1$ during the free precession period of length $T$, and $h(t)=\sin(\Omega_R t)$ ($h(t)=-\sin(\Omega_R t)$) during the first (second) $\pi/2$ pulse, where $\Omega_R$ is the Rabi frequency. The total measurement consists of $n$ repetitions of Ramsey interrogations. Each interrogation cycle is of duration $T_c=T+T_D+2t_p$, where $T_D$ is the dead time, $r=T/T_c$ is the duty cycle, $t_p$ the pulse duration and $\tau=n T_c$ is the total measurement time. For $\pi/2$ pulses, $t_p = \pi/(2 \Omega_R)$, the noise transfer function is then
\begin{equation}\label{eq:NoiseTrans}
\begin{split}
|\tilde{H}(f)|^2 & = \frac{1}{n^2 T^2}\frac{\Omega_R^2}{((2 \pi f)^2 - \Omega_R^2)^2}  \left(  \frac{\Omega_R}{\pi f} \sin(\pi f T) \right. \\
& \left. + 2 \cos \left(\pi f T + \pi^2 \frac{f}{\Omega_R} \right) \right)^2 \frac{\sin^2(\pi f n T_c)}{\sin^2(\pi f T_c)} .
\end{split}
\end{equation}
\noindent Here, the last term captures the finite dead time in-between measurements, which can significantly alter the scaling of the noise with averaging time. We therefore consider two cases, that of zero dead time ($T_D=0$), and that of finite dead time ($T_D>0$). The transfer functions for three representative cases are plotted in Fig.~\ref{Fig-TransNoise}.

\begin{figure}
\centering
\includegraphics[width=0.9\columnwidth]{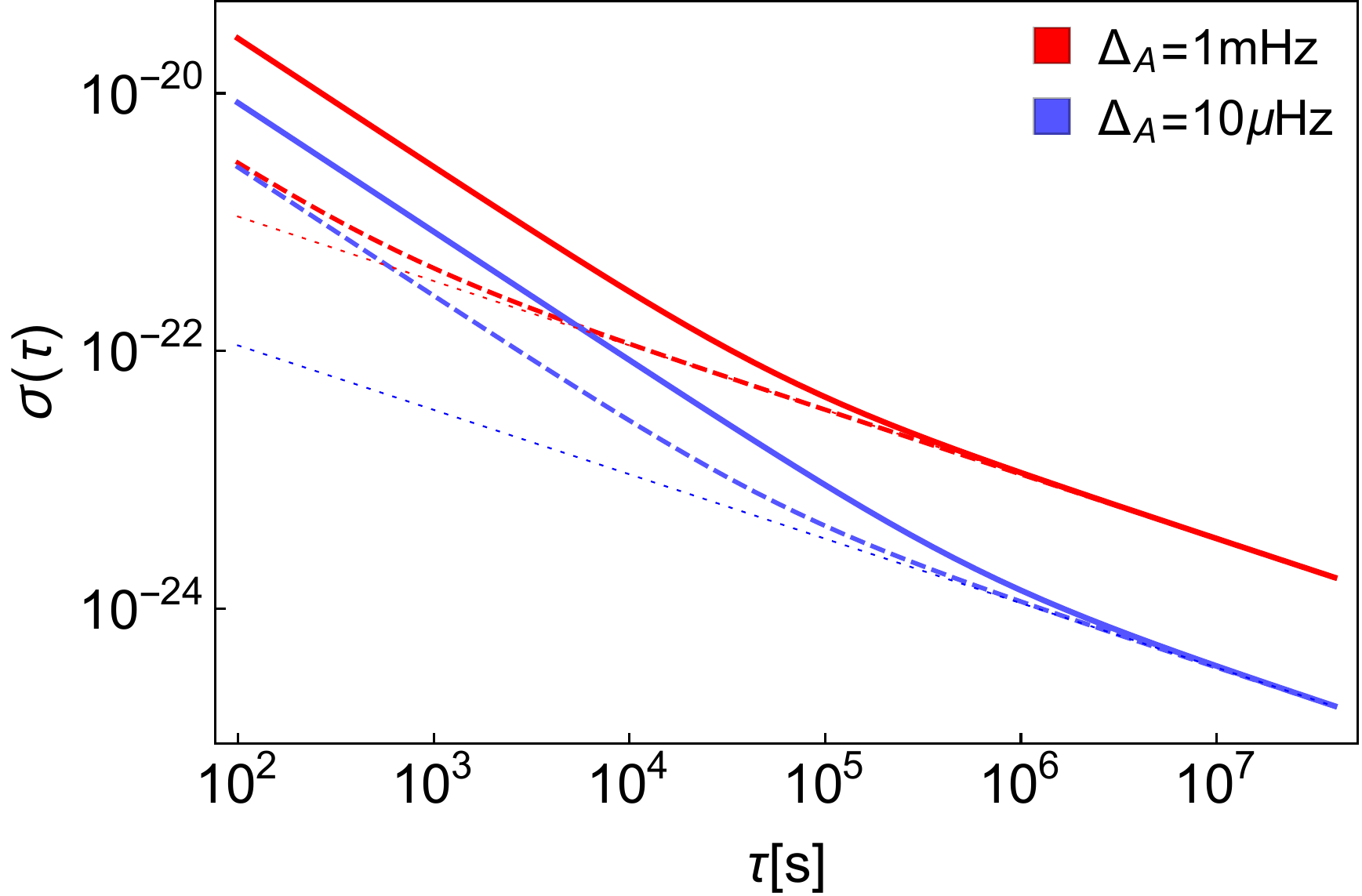}
\caption{\small Fractional frequency instability as a function of averaging time $\tau$ as given in Eq.~\eqref{eq:var2}, for the case of zero dead time ($r=1$). The red curves correspond to an atomic linewidth $\Delta_A=1$~mHz as in the main text, while the blue curves are for a narrower atomic transition with $\Delta_A = 10 \mu$Hz. The thick and dashed lines differ by the received optical power: $95$~pW (red thick line), $10$~nW (blue thick line) and $1 \mu$W (blue and red dashed lines). The dotted lines show the atom projection noise limit as given in Eq.~2 of the main text. For short averaging times photon shot noise dominates and the noise scales with $1/\tau$, while at long averaging times atom projection noise dominates and the noise scales as $1/\sqrt{\tau}$. }
\label{Fig-tau}
\end{figure}

If there is no dead time and for $\Omega_R\gg B$, the noise transfer function in Eq.~\ref{eq:NoiseTrans} simplifies dramatically to become $|\tilde{H}(f)|^2=\textrm{sinc}^2(\pi f T)$ and the integral can be computed analytically to give $\int_0^{\infty}df \textrm{sin}^2(\pi f \tau)/((2 \pi f)^2+B^2) = (1-e^{-B \tau})/(8 B)$. Including the atom projection noise given in the main text, the overall variance in frequency measurement for $r=1$ is therefore
\begin{equation}\label{eq:var}
\sigma^2 = \frac{1}{(2 \pi \nu)^2 T \tau} \left( \frac{1}{N} + \frac{1-e^{-B \tau}}{2 \tau / T} \left( \frac{\Delta_L}{B} + \frac{h \nu}{\eta P_B} B \right) \right) .
\end{equation}
\noindent For optimized loop bandwidth $B_{opt}$ and in the limit $B \tau \gg 1$, the above expression becomes
\begin{equation}\label{eq:var2}
\sigma^2 = \frac{1}{(2 \pi \nu)^2 T \tau} \left( \frac{1}{N} + \frac{T}{ \tau} \sqrt{\frac{h \nu \Delta_L}{\eta P_B}} \right) .
\end{equation}
\noindent In this limit, the contribution from laser phase noise averages down as $\sigma_L^2\propto1/\tau^2$, consistent with other Doppler tracking detectors \cite{DopplerReview}. As a result, the photon shot noise averages down faster than the atom projection noise, and at long averaging times atom projection noise will dominate (see Fig.~\ref{Fig-tau}).

Zero dead time clock operation has been realized using interleaved measurements of two clocks \cite{KasevichTwoClocks,LudlowTwoClocks,Meunier}. However, if our detector is restricted to only a single clock per satellite, detector operation will likely include a small but finite time between subsequent measurements, which introduces additional noise through a process known as the Dick effect \cite{DickEffect}. For the clock GW detector, the dead time results in aliasing down of the high frequency noise in the PLL, resulting in differential frequency noise in the two-clock comparison, which can limit the differential clock stability. We emphasize that this differential Dick noise is distinct from the aliased laser frequency noise traditionally referred to as Dick noise. While ``traditional" Dick noise will also be present in each individual clock making up the detector, it is common mode and will be cancelled out in the synchronous comparison. In order to account for the differential Dick noise due to finite dead time ($T_D>0$), integration over the full transfer function has to be performed. This was done using numeric integration for a finite number of measurements $n$, and analytically for the limit $n \rightarrow \infty$.

Any finite dead time will alias the high frequency differential laser noise in the PLL into differential white frequency noise, resulting in a Dick noise term which scales as $\sigma_D^2\propto1/\tau$. Because this term averages down more slowly than the $\sigma_L^2\propto1/\tau^2$ term in Eq.~\eqref{eq:var2}, at some finite number of measurements, $n_D$, $\sigma_D$ will begin to dominate over $\sigma_L$. Numerical integration of Eq.~\eqref{eq:VarLight2}, with Eqs.~\eqref{eq:PLLPSD} and \eqref{eq:NoiseTrans} for finite $n$, and in the limits $\Omega_R\gg B_{opt}$, $T_D\gg1/\Omega_R,$ and $T\gg T_D$, yields $\sigma_D^2\approx (n/n_D)\times\sigma_L^2$, where $n_D\approx1/(2 \pi T_D B_{opt})$. Therefore, for $n$ sequential measurements the differential Dick noise can be safely ignored for small enough dead times, with the condition $nT_D\ll 1/(B_{opt})$, while in the limit of many measurements, the Dick noise will always dominate.

For current individual optical lattice clocks $T_D\approx1$~s, and $T_D\gg 1/B_{opt}.$ In this case, and in the limit of many measurements, the full integral in Eq.~\eqref{eq:VarLight2} with Eqs.~\eqref{eq:PLLPSD} and \eqref{eq:NoiseTrans} can be evaluated analytically using the property of the Fej\'{e}r-kernel: $F(x)=\sin^2(n x)/(\sin^2(x) n) \rightarrow \pi \delta(x)$ on  $x \in[-\pi/2,\pi/2]$. The resulting frequency uncertainty from laser noise becomes

\begin{equation}\label{eq:var3}
\begin{split}
\sigma^2_L & = \frac{8 \Omega_R^2 B}{(2 \pi \nu)^2 \tau T^2 } \sum_{k=0}^{\infty} \frac{\frac{\Delta_L}{B} + \frac{h \nu }{\eta P_B}B}{\left(2 \pi \frac{k r}{T} \right)^2 + B^2} \frac{1}{\left( \left(2 \pi \frac{k r}{T} \right)^2 - \Omega_R^2\right)^2} \\
& \times  \left(\Omega_R \sin(\pi k r) + 2 \pi \frac{k r}{T} \cos(\pi k r + 2 \pi t_p k r/T)  \right)^2.
\end{split}
\end{equation}

The contribution from laser phase noise now averages down more slowly, $\sigma_L^2\propto1/\tau$, thereby competing directly with atom projection noise. However, the Rabi frequency $\Omega_R$ used in the Ramsey sequence can be used as a lowpass filter on the atomic response in order to limit the susceptibility to high frequency noise resulting from the Dick effect, as shown in Fig.~\ref{Fig-TransNoise}. As long as the PLL bandwidth $B_{opt}$ is kept above $\Omega_R$, the PLL can bridge the dead time between subsequent Ramsey sequences, suppressing the differential laser noise, and the noise spectrum experienced by the atoms is simply the photon shot noise from the PLL detection during the Ramsey control pulses. For $\Omega_R\ll B_{opt}$, Eq.~\ref{eq:var3} then simplifies to (now again including atom projection noise)

\begin{equation}\label{eq:var6}
\sigma^2 = \frac{1}{(2 \pi \nu)^2 T \tau } \left( \frac{1}{N} + \frac{2}{r}  \frac{h \nu}{ \eta P_B}\Omega_{R} \right)
\end{equation}

This corresponds to the intuitive condition that the number of photons received at satellite B during the Ramsey control pulses must be larger than the number of atoms $N$ used in each run of the measurement.

\section{Time constraints for narrowband signal observation}\label{App:Time}
The narrowband nature of the clock detector means that averaging and observation time will be fundamentally limited by the duration of the GW at the specific frequency of interest. Compact binary inspirals produce continuous GWs which experience a chirp towards higher frequencies, given by \cite{maggiore2008gravitational,Harms}:
\begin{equation}
\label{eq:chirp}
\dot{f}= \frac{96}{5} \pi \left( \frac{\pi G M_c}{c^3}\right)^{5/3} f^{11/3},
\end{equation}\\
where $G$ is the gravitational constant and $M_c = (m_1 m_2)^{3/5}/(m_1+m_2)^{1/5} $ is the effective chirp mass of a binary system with masses $m_1$ and $m_2$.
The number of GW cycles in a time $t \in [t_1,t_2]$ small compared to the GW period is $dn_{cyc}=fdt$, or $n_{cyc}=\int_{f_1}^{f_2}df f/\dot{f}$. Assuming $f_2-f_1 \approx 1/T_{max}$, we find that the time the GW is within this frequency range is given by
\begin{equation}\label{eq:tauGW}
\tau_{\text{GW}}=\frac{n_{cyc}}{f} \approx 2.5 \times 10^{10} \textrm{s} \left( \frac{10 \textrm{mHz}}{f}\right)^{8/3} \left( \frac{2.6 M_{\odot}}{M_c}\right)^{5/3} ,
\end{equation}
\noindent where $M_{\odot}=2\times 10^{30}$~kg is the solar mass, and we have normalized $M_c$ to the mass value for an inspiral of two objects with  $m_1=m_2=3 M_{\odot}$. For such sources, and for frequencies in the $\sim$10 mHz range, the GW has an essentially fixed frequency over hundreds of years. For heavier sources, however, $\tau_{\text{GW}}$ can be much shorter; for a black hole binary as detected by LIGO ($m_1=36 M_{\odot}, m_2=29 M_{\odot}$) we have $\tau_{\text{GW}} \approx 15$~years in the above frequency range around $f=10$~mHz.
For optimal GW detection, we require $\tau_{av}<\tau_{\text{GW}}$, which is reasonable for most sources expected in the frequency range of interest. We also note that this is not a strict limitation for a source with a known frequency chirp, as the measurement sequence can be easily adapted to chirp the detection window along with the source.

\bibliography{GWclockBib}

\end{document}